\documentclass[twocolumn,showpacs,prl]{revtex4}
\usepackage{amsmath,amsthm,amscd}
\usepackage{dcolumn}

\usepackage{graphicx} \usepackage{epsfig} \begin{document}
\title{Spin swap vs. double occupancy in quantum gates}
\author{T. A. Kaplan and C. Piermarocchi}
\affiliation{Department of Physics and Astronomy and Institute for
Quantum Sciences, \\ Michigan State University, East Lansing, Michigan
48824}

\begin{abstract} We propose an approach to realize quantum
gates with electron spins localized in a semiconductor that uses double occupancy to
advantage. With a fast (non-adiabatic) time control of the tunnelling, the probability
of double occupancy is first increased and then brought back exactly to zero. The
quantum phase built in this process can be exploited to realize fast quantum
operations. We illustrate the idea focusing on the half-swap operation, which is the
key two-qubit operation needed to build a CNOT gate.  \end{abstract}

\pacs{75.10.Pq}

\date{\today}
\maketitle

Many solid state proposals for quantum computation make use of the
spin of an electron localized in a quantum dot~\cite{friesen03,
loss98} or by a donor ion~\cite{vrijen00} as a qubit. In these schemes
the overlap between the electron wave functions in neighboring dots is
controlled by gates, and the exchange interaction provides the
mechanism for the realization of conditional operations on two
qubits. Among the requirements for such proposals is that one wants no
or small double occupancy, i.e. small probability to have the two
electrons in the same dot or donor. In fact, the state space
$\mathcal{S}$ for 2 qubits has to be 4-dimensional and the mixing with
the states of double occupancy brings the physical system out of the
computational space.  Hu and Das Sarma~\cite{hu} observed that the
double occupancy will in principle degrade any implementation of
quantum gates, and noted the intrinsic conflict of trying to minimize
this double occupancy while making the overlap large enough to realize
a sizeable two-qubit operation, like e.g. a spin swap. The problem was
further analyzed in Ref.~\cite{schliemann}, where it was shown that
using an adiabatic control in the quantum dots case, the double
occupancy was small enough so that the degradation can be made
negligible, consistently with the requirement of large enough
overlap. The operation critical to quantum computing is the
\emph{half-swap} operation (also written as $\sqrt{SWAP}$),
i.e. having the state going half of the way to the complete swap, and
we focus on this process to illustrate the approach.

We show here that \emph{a half-swap operation without degradation can be realized by
designing a time control where the probability of double occupancy first increases and
then is brought back exactly to zero. In this cycle the quantum state in $\mathcal{S}$
picks up a phase that is exploited in completing the process.} As a consequence, the
swap can be much faster than that in the case where double occupancy is kept small at
all times. We will consider explicitly two cases where the control of the tunneling
amplitude follows in time a square and a hyperbolic secant pulse shape. The conditions
for a perfect $\sqrt{SWAP}$, i. e. a fast half-swap gate with double occupancy
identically zero, can be given in these two cases by simple analytic rules.  The
results are exact within the single band model considered, the validity of which will
be discussed below for realistic systems.


The dynamics during the half swap can be described by using two spatial orbitals,
which we call $w_a$ and $w_b$ ($w$ for Wannier functions), orthonormal, and localized
respectively around the two sites $a$ and $b$~\cite{hu, schliemann, burkard}. The
$w$'s are linear combinations of non-orthogonal ``atomic" functions whose overlap is
$\Lambda$. This implies a 4-state one-electron basis, and therefore a 6-dimensional
complete vector space for 2 electrons.~\cite{slater} The Hamiltonian $H$ contains the
kinetic and potential energies, plus a magnetic field interacting with the orbital
magnetic moment. The quantum control is realized by changing the profile of the
potential energy, therefore modifying the tunnelling amplitude between the two sites.
We will not consider a magnetic field term, since it is irrelevant in the proposed
scheme. The 2-electron basis used is
\begin{subequations}
\label{5}
\begin{eqnarray}
|\Phi_1\rangle&=&2^{-1/2}(c^\dagger_{a\uparrow}c^\dagger_{b\downarrow}
-c^\dagger_{a\downarrow}c^\dagger_{b\uparrow})|0\rangle  \\
|\Phi_2\rangle&=&2^{-1/2}(c^\dagger_{a\uparrow}c^\dagger_{a\downarrow}
+c^\dagger_{b\uparrow}c^\dagger_{b\downarrow})|0\rangle \\
|\Phi_3\rangle
&=&2^{-1/2}(c^\dagger_{a\uparrow}c^\dagger_{a\downarrow}
-c^\dagger_{b\uparrow}c^\dagger_{b\downarrow})|0\rangle \\
|\Phi_4\rangle
&=&c^\dagger_{a\uparrow}c^\dagger_{b\uparrow}|0\rangle \\
|\Phi_5\rangle
&=&2^{-1/2}(c^\dagger_{a\uparrow}c^\dagger_{b\downarrow}
+c^\dagger_{a\downarrow}c^\dagger_{b\uparrow})|0\rangle \\
|\Phi_6\rangle&=&c^\dagger_{a\downarrow}c^\dagger_{b\downarrow}|0\rangle~,
\end{eqnarray}
\end{subequations}
where $c^\dagger_{\nu\sigma}$ creates an electron in the orbital $w_{\nu}$ with spin
$\sigma$.  Physically we speak of $\Phi_{\nu}$ with $\nu =1,4,5,6$ as having 1
electron near each site, while for $\nu=2,3$, the states have 2 electrons on one site,
and are called doubly-occupied states. Assuming inversion symmetry, the time-dependent
Hamiltonian $H_{ij}(t)\equiv \langle\Phi_i|H(t)|\Phi_j\rangle$ reduces by symmetry to
four $1 \times 1$'s (the triplet $H_{44}$, $H_{55}$, and $H_{66}$, and the {\it
ungerade} singlet $H_{33}$), plus a $2 \times 2 $ which yields the remaining two
(\emph{gerade}) singlets.  We assume that before applying the control pulse, at time
$t=-\infty$, the wave function is $|\Psi(-\infty)\rangle
=c^\dagger_{a\uparrow}c^\dagger_{b\downarrow}|0\rangle
=2^{-1/2}(|\Phi_1\rangle+|\Phi_5\rangle)~.$ Given the form of the initial state and
the symmetry properties discussed above, $|\Phi_3 \rangle$, $|\Phi_4\rangle$ and
$|\Phi_6\rangle$ do not enter in the dynamics and we calculate
$|\Psi(t)\rangle=a(t)|\Phi_1\rangle+b(t)|\Phi_2\rangle +c(t) |\Phi_5\rangle $ by the
(time-dependent) Schr\"{o}dinger equation
\begin{equation} i\frac{d}{dt}\left[
\begin{array}{c} a(t)\\ b(t)\\ c(t) \end{array}\right]^T=\left[
\begin{array}{ccc} 0 & \frac{\Omega(t)}{2}& 0 \\ \frac{\Omega(t)}{2}&
\Delta &0 \\ 0 & 0& J_P(t) \end{array} \right] \left[ \begin{array}{c} a(t)\\ b(t)\\
c(t) \end{array}\right] \end{equation} where $\hbar\Omega(t)=2 H_{12}(t)$,
$\hbar\Delta=H_{22}-H_{11}$ and $\hbar J_P(t)=H_{55}(t)-H_{11}(t)$. Notice that
$\Omega(t)$ is first order in the overlap $\Lambda$, while $\Delta$ depends only on
the on-site and long-range Coulomb interaction: $\hbar\Delta =\langle
w_a,w_a|v|w_a,w_a\rangle - \langle w_a,w_b|v|w_a,w_b\rangle $, where $v$ is the
Coulomb interaction between the two electrons. This term is of order $\Lambda^0$, and
therefore only weakly dependent on the tunnelling amplitude. So we assume $\Delta$
constant during the gate operation. $\hbar J_P(t)=-2 \langle w_a,w_b|v|w_b,w_a\rangle$
is the potential exchange~\cite{anderson}, which is O($\Lambda^2)$. The hopping term
$\hbar \Omega(t)=4\langle w_a|h_1|w_b\rangle+4\langle w_a,w_b|v|w_a,w_a\rangle$, where
$h_1$ is the single particle kinetic and potential energy, is responsible for the
kinetic exchange. Both $J_P(t)$ and $\Omega(t)$ are controlled by the tunnelling
amplitude through their dependence on the overlap $\Lambda$. We remark that these
parameters simplify in the Hubbard model to $J_P(t)=0$ and $\hbar \Delta=U$. In this
case the hopping amplitude $\Omega(t)$ is the only mechanism for the exchange
interaction ($J_K \cong -\hbar^2\Omega(t)^2/2U$ for small $\Lambda$, the kinetic
exchange~\cite{anderson}). The presence of the potential exchange $J_P(t)$ is not
essential in the realization of the proposed scheme, but we keep it in order to prove
the reliability of the scheme in the most general case.

It is now simple to calculate the quantities we want to monitor. One is the
expectation value of $s_z=n_{a\uparrow}-n_{a\downarrow}$, the z-component of spin at
site $a$. ($n_c$ are occupation numbers.) It is sufficient to consider just one site
since the total z-component of spin is conserved and so must remain 0. Using the
property $s_z|\Phi_1\rangle=|\Phi_5\rangle$, $s_z|\Phi_5\rangle=|\Phi_1\rangle$, and
$s_z|\Phi_2\rangle=0$, we find
\begin{equation}
\sigma_z(t)\equiv \langle\Psi(t)|s_z|\Psi(t)\rangle=2 Re[a(t)c^*(t)]~. \label{sigmaz}
\end{equation}
We also want the probability $P_d$ of double occupancy, i.e. the
probability of finding the system in the state $\Phi_2$. This is
\begin{equation}
P_d(t)=|\langle\Phi_2|\Psi(t)\rangle|^2= |b(t)|^2 \label{12}~.
\end{equation}
In order to have the perfect $\sqrt{SWAP}$ at the end of the pulse,
i.e. at $t=+\infty$, we need to design the control in such a way that
the two conditions
\begin{subequations}
\label{c1c2}
\begin{eqnarray}
P_d(+\infty)=0 \label{c2} \\
\sigma_z(+\infty)=0 \label{c1}
\end{eqnarray}
\end{subequations}
are verified. One can see that Eq.~(\ref{c1c2}) implies
$|\Psi(+\infty)\rangle=2^{-1/2}(c^\dagger_{a\uparrow}c^\dagger_{b\downarrow} \pm i
c^\dagger_{a\downarrow}c^\dagger_{b\uparrow})|0\rangle$ (Bell entangled states). In
the usual approach a weaker constraint than the one in Eq.~(\ref{c2}) is used (i.e.
$P_d(+\infty)\ll 1$) satisfied by slow adiabatic switching and small mixing with
doubly occupied states. This condition is implicit when we describe the system of two
localized electrons using the (effective) Heisenberg Hamiltonian. Our approach
consists in finding first the general conditions on the control to satisfy exactly
Eq.~(\ref{c2}), which does not imply necessarily adiabaticity and small average mixing
with the doubly occupied states, and then adjust the remaining control parameters to
satisfy Eq.~(\ref{c1}).

The time dependence of $c(t)$ can be integrated directly from the
Schroedinger equation and we obtain
\begin{equation}
c(+\infty)=2^{-1/2} \exp(-i\int_{-\infty}^{+\infty}J_P(t)~dt)~.
\end{equation}
The dynamics of $a(t)$ and $b(t)$ can be mapped to the evolution of a pseudospin in
the presence of a time dependent effective magnetic field. The condition in
Eq.~(\ref{c2}) means that the pseudospin makes a closed loop in the Bloch sphere,
while the condition in Eq.~(\ref{c1}) ensures that the phase picked up by the
pseudospin evolution during such a closed loop compensates the phase due to the
potential exchange for a perfect $\sqrt{SWAP}$.  The analytic solution for the
evolution of a spin in a time dependent magnetic field can be found only in particular
cases~\cite{ishk00}, and we will discuss here two of them: the Rabi
solution~\cite{allen}, corresponding to a square pulse shape, and the Rosen and Zener
solution~\cite{rosen32} corresponding to a hyperbolic secant pulse shape.

The Rabi solution describes the precession of the pseudospin in a constant effective
magnetic field. A square pulse in the tunnelling amplitude keeps this effective field
on for a time $T$, and then turns it off. We can write the hopping term as a constant
$\Omega(t)=\Omega$, and consider the time evolution given by the Rabi solution from a
time $t=0$ to $t=T$ (just the solution for $t$-independent $H$). In a similar way the
potential exchange term $J_P(t)=J_P$ for $0 \le t \le T$ and zero otherwise. The
probability of double occupancy at the end of the pulse is given by:
\begin{equation}
P_d(T)=|b(T)|^2=\frac{\Omega^2/2}{\Omega^2+\Delta^2}
\sin^2\left(\frac{\sqrt{\Omega^2+\Delta^2}}{2}T\right)~.
\end{equation}
Therefore the condition in Eq.~(\ref{c2}) is satisfied by fixing the
length of the pulse to
\begin{equation}
T_n= \frac{2 n \pi}{\sqrt{\Omega^2+\Delta^2}}~.
\end{equation}
The integer $n$ identifies the number of full precessions of the
pseudospin during the pulse.  We find from the Rabi solution that $
a(T_n)=\exp (-i\Delta T_n /2)/\sqrt{2}$. Using this solution in
Eq.~(\ref{sigmaz}), the condition in Eq. (\ref{c1}) for the perfect
$\sqrt{SWAP}$ is rewritten as
\begin{equation}
\cos n\pi\left(\frac{\Delta-2J_P}{\sqrt{\Omega^2+\Delta^2}} \right)=0. \label{genR}
\end{equation}
This can be recast in terms of the separation between the triplet and
the lower singlet eigenvalues $J$, and the separation between the two
\emph{gerade} singlets $\Delta E$ as
\begin{equation}
\frac{2 n J}{\Delta E}=\frac{2k+1}{2}~,
\label{condR}
\end{equation}
where $k$ is any integer.

\begin{figure}
\centering \includegraphics[height=8. cm, angle=270]{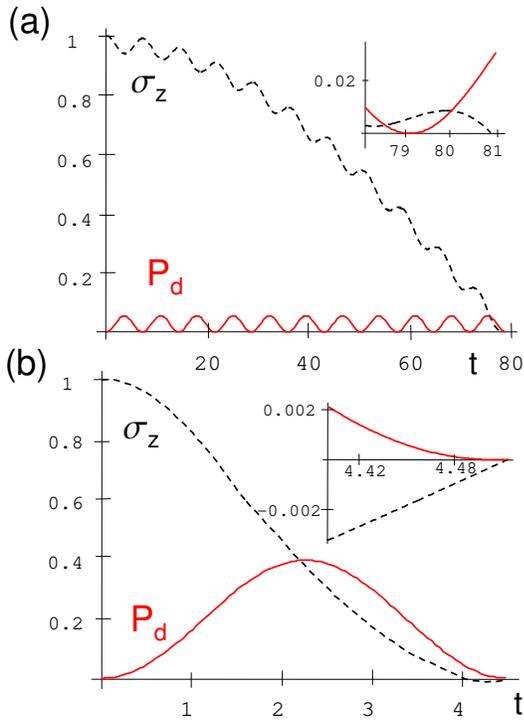}
\caption{$\sigma_z(t)$ and $P_d(t)$ for two electrons confined by
 hydrogenic potentials. Infinite tunnelling barriers are switched off
 during the time interval $0<t<T$. Time $t$ is in units of
 $\hbar$/Ry$^*$. (a) The separation between the sites is $4.53$ au*. At
 the end of the pulse we have a half swap with a small but nonzero
 $P_d$. (b) The separation is $2.38$ au* and the energy levels satisfy
 the condition in Eq.~(\ref{condR}). The half swap is realized in a
 shorter time and with $P_d=0$ identically at the end of the pulse
 (see inset).}
\label{fig:sigma}
\end{figure}

To illustrate the idea we show in Fig.~\ref{fig:sigma} the evolution of $\sigma(t)$
and $P_d(t)$ in the square pulse case calculated assuming a hydrogenic confinement
potential and localized functions of the Slater form, which models well the case of
electrons confined in shallow donors. We want to stress that the general approach we
are suggesting is independent of this particular choice. In fact, for any reasonable
form of the confinement potential it is possible to modify the ratio in
Eq.~(\ref{condR}) by changing the maximum tunnelling amplitude or the separation
between the dots/impurities. When the separation is large, so the overlap
is small (or the tunnelling barrier is large), there is a small admixture of
doubly-occupied states at any time. This is the region where the low-lying states are
described accurately by the Heisenberg Hamiltonian $2J\mathbf{S}_a\cdot\mathbf{S}_b$.
The spins here are associated with the sites; e.g. $S_a^+=c^\dagger_{a\uparrow}
c_{a\downarrow}$~\cite{anderson}. This case is considered in Fig.~\ref{fig:sigma}~(a),
obtained with a separation between the localized electrons of 4.53 au$^*$. We indicate
by au$^*$ and Ry$^*$ the atomic units scaled by the electron effective mass and the
static dielectric constant. Time is in units of $\hbar$/Ry$^*$. For a typical GaAs
based system 1 Ry$^*$ $\sim$ 5 meV, 1 au$^*$ $ \sim $ 80 \AA, and $\hbar/Ry* \sim $
0.1 ps.  We see that at the end of the pulse we satisfy the condition for the half
swap $\sigma_z=0$ with a small double occupancy $P_d \sim 0.03 $ (see inset in
Fig.~1~(a)). The new approach is illustrated in Fig.~\ref{fig:sigma}~(b) corresponding
to a separation of 2.38 au$^*$. At this separation the energy levels satisfy the
condition in Eq.~(\ref{condR}) yielding a perfect half swap with both $\sigma_z=0$ and
$P_d=0$ identically. Fig.~\ref{fig:sigma}~(b) shows a marked increase in swap speed
and a much higher probability of double occupancy during the gate compared to
Fig.~\ref{fig:sigma}~(a). The case in Fig.~\ref{fig:sigma}~(b) corresponds to $n=1$,
higher $n$ will give a perfect half swap in longer times.

The use of square pulses gives a direct illustration of the new approach. However, for
practical purposes, a continuous time dependence of the pulse is desirable. Using the
picture of the pseudospin in an effective magnetic field, we need to design in the
general case a $\mathbf{B}_{eff}(t)$ that generates closed loops in the Bloch sphere.
Once this first condition is realized, the condition on the control parameters for a
perfect half swap can be found accordingly. This is possible for the hopping amplitude
of the form $\Omega(t)=\Omega/\cosh\left( \pi t/T\right)$. Then
$\int_{-\infty}^{+\infty}\Omega(t)~dt=\Omega T$, the same as for the square pulse. The
probability of double occupancy at the end of the pulse (at $t=+\infty$) can be
obtained from the Rosen and Zener result~\cite{rosen32}
\begin{equation}
P_d=\frac{\sin^2\Omega T/2}{2 \cosh^2\Delta T/2}.
\label{rzres}
\end{equation}
In the adiabatic approach the condition $\Delta T\gg 1$ in the denominator makes this
probability small.  Our approach takes advantage of the numerator in
Eq.~({\ref{rzres}), which implies that for
\begin{equation}
T_n= \frac{2 n \pi}{\Omega}
\end{equation}
we will have no double occupancy at the end of the pulse. In order to
determine the phase of $a(+\infty)$, we use the transformation
$2z=\tanh(\pi t/T)+1$ to map the infinite time interval to $[0,1]$,
and the equation for the coefficient $a(z)$ reads
\begin{equation}
(z^2-z)a^{\prime \prime}+\left(\frac{1}{2}+\frac{i\Delta T}{2 \pi}-z\right)a^{\prime}
+ \left(\frac{\Omega T}{2\pi}\right)^2 a=0~.
\label{hyper}
\end{equation}
This is the hypergeometric equation and in general we obtain a
solution in terms of a linear combination of hypergeometric
functions. However, it is interesting to notice that if the condition
for no double occupancy is met, then $\left(\Omega
T/2\pi\right)=n$, and we will fall in the degenerate case of the
hypergeometric equation. As discussed in Ref.~\cite{erdelyi53}, the
solution in this case consists of a polynomial of degree $n$. We find
that the solution to Eq.~(\ref{hyper}) with the initial condition
$a(0)=1$ and with $|a(1)|^2=1$ can be written explicitly as
$a_n(z)=n!P^{\alpha_n,\alpha_n^*}_{n}(1-2z)/(\alpha_n+1)_n ~,$ where
$P^{\beta,\gamma}_{n}(x)$ are Jacobi polynomials, $\alpha_n= -1/2 + i n
\Delta/\Omega$, and
$(x)_n=\Gamma\left(x+n\right)/\Gamma\left(x\right)$.  The condition in
Eq.~(\ref{c1}) can then be written in a compact form using the
hypergeometric function (which in this case terminates after $n$ terms
in the series) as
\begin{equation}
Re[F(-n,n,1/2+i n \Delta/\Omega,1) e^{i 4 n J_P/\Omega}]=0~,
\label{condRZ}
\end{equation}
which is the parallel of Eq.~(\ref{genR}) for this continuous pulse shape. The fact
that the potential exchange $J_P(t)$ is O($\Lambda^2)$ led us to take $J_P(t)=J_P
\mbox{sech}^2(t/T)$ in Eq.~(\ref{condRZ}), although the solvability and essential
aspects of the solution are not limited to this. We show in Fig.~\ref{fig2} the
realization of a perfect half swap in the hydrogenic case. At the separation of about
2.36 au* the matrix elements of the Hamiltonian satisfy the condition in
Eq.~({\ref{condRZ}) and a perfect half swap can be realized. The case in
Fig.~\ref{fig2} corresponds to $n=1$ and we have checked that for higher $n$ we can
always find at least one separation to satisfy Eq.~(\ref{condRZ}). The inset shows the
exponential converging of $P_d$ and $\sigma_z$ toward the desired value. We remark
that in GaAs based systems this translates in a half swap operated with an accuracy of
$10^{-10}$ in a time of the order of 3 ps.
\begin{figure}
\centering \includegraphics[scale=0.3]{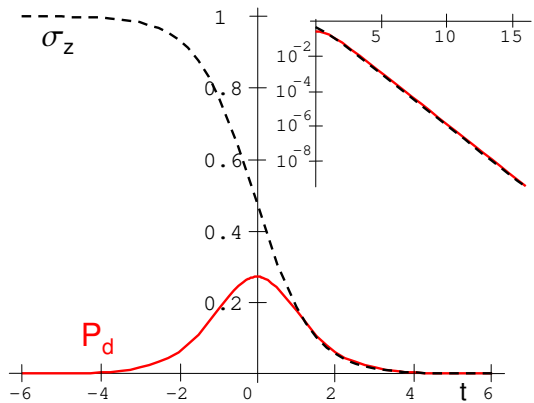} \caption{$\sigma_z(t)$
and $P_d(t)$ for two electrons confined by hydrogenic potentials with barriers
controlled using a hyperbolic secant pulse shape.  The separation is 2.36$\cdots$
au$^*$. Time $t$ is in units of $\hbar$/Ry$^*$. The inset shows the Log of $\sigma(t)$
and $P_d(t)$ at longer times.} \label{fig2}
\end{figure}

We now consider likely sources of error and other difficulties to be
expected in a real implementation of the ideas proposed. We pose the
question, if $\Omega$ and $\Delta$ deviate slightly from values needed
for a perfect half-swap, how much do $\sigma(+\infty)$ and
$P_d(+\infty)$ differ from their ideal values of zero? Using the
analytic solutions we proved that if the deviation of the parameters
is O($\epsilon$) then the deviation $\sigma(+\infty)=$O$(\epsilon)$
and $P_d(+\infty)=$O$(\epsilon^2)$.
Higher intra-dot excitations will complicate our considerations. If they are large
compared to $\hbar \Delta$, however, their effect will be small. To get an estimate of
the requirements, consider these excitations for the infinite square well, side length
$L$. Then $\delta E=3\pi^2\hbar^2/2m^*L^2=6\pi^2(\text{au*}/L)^2 \text{Ry*}$, is the
first excitation energy, and $\hbar \Delta\approx 2 e^2/(\epsilon
L)=2(\text{au*}/L)\text{Ry*}$, so $\delta E/\hbar \Delta\approx 3\pi^2(\text{au*}/L)$.
For two GaAs quantum dots of lateral size $L=100 \AA$, this gives $\delta
E/\hbar\Delta\approx 27$ and $\hbar \Delta\approx 2\text{Ry*}$, satisfying the
requirement $\delta E/\hbar \Delta>>1$. Finally, we remark that in order to take
advantage of the proposed scheme the time resolution needed for controlling the gate
is related to the intra-dot Coulomb repulsion $\hbar \Delta$. In principle, a
subpicosecond switching time for a pulse gate is not out of the
question~\cite{tessmer}.  By increasing the dot size $L$ while maintaining a large
$\delta E/\hbar\Delta $ it is possible to increase the resolution time $\Delta^{-1}$.
We estimate that this condition can be well satisfied for large ($\sim$ 300-400 \AA)
GaAs quantum dots in such a way to limit the resolution to the picosecond range, which
is within state-of-the-art capabilities.

In conclusion, we have shown that by non-adiabatic control it is possible to take
advantage of the double occupied states to make fast and accurate half swap
operations.  The idea can be applied to many systems currently under investigation for
experimental realization of quantum computing. Even in systems where the mixing with
double occupancy is small, the simple rules given here can be used as an error
correction method. Possible generalizations to more than two spins~\cite{lidar04} are
under consideration. It is also of interest to consider more general pulse shapes,
particularly with the object of seeing what forms of $\mathbf{B}_{eff}(t)$ will allow
the pseudospin to return to its starting value (i.e. corresponding to $P_d=0$).


\begin{acknowledgements}
We thank S. H. Tessmer, M. I. Dykman, and S. D. Mahanti for
enlightening discussions. C.P. acknowledges support by NSF under
contract DMR-0312491.
\end{acknowledgements}

\end{document}